\documentclass[twocolumn,showpacs,showkeys,amsmath,amssymb,aps,pre]
{revtex4-1}
\usepackage{float}
\usepackage[final]{graphicx} 
\usepackage{dcolumn}
\usepackage{bm}
\usepackage[dvips]{color}

\newcommand\Rey{\mbox{\text{Re}}}  

\newcommand\eg{e.g.\ }
\newcommand\ie{i.e.\ }

\begin{document}

\title{On the nature of laminar-turbulence intermittency in shear flows}

\author{M. Avila$^{1,2}$ and B. Hof$^{2,3}$}

\affiliation{$^{1}$Institute of Fluid Mechanics,
  Friedrich-Alexander-Universit\"at Erlangen-N\"urnberg, 91058
  Erlangen, Germany \\$^{2}$Max Planck Institute for Dynamics and
  Self-Organization (MPIDS), 37077 G\"ottingen,
  Germany\\ $^3$Institute of Science and Technology Austria, 3400
  Klosterneuburg, Austria}

\date{\today}

\begin{abstract}
  In pipe, channel and boundary layer flows turbulence first occurs
  intermittently in space and time: at moderate Reynolds numbers
  domains of disordered turbulent motion are separated by quiescent
  laminar regions. Based on direct numerical simulations of pipe flow
  we here argue that the spatial intermittency has its origin in a
  nearest neighbor interaction between turbulent regions. We further
  show that in this regime turbulent flows are intrinsically
  intermittent with a well defined equilibrium turbulent fraction but
  without ever assuming a steady pattern. This transition scenario is
  analogous to that found in simple models such as coupled map
  lattices. The scaling observed implies that laminar intermissions of
  the turbulent flow will persist to arbitrarily large Reynolds
  numbers.
\end{abstract}

\maketitle 

\section{Introduction} 

In fluid flow inertia has in general a destabilizing effect, whereas
viscous forces tend to quickly restore smooth motion. The balance
between the two is expressed in the dimensionless Reynolds number
$\Rey= L U / \nu$, where $L$ and $U$ are a characteristic length and
velocity, respectively, and $\nu$ the kinematic viscosity of the
fluid. As \Rey\ is increased laminar flows tend to be unstable, and in
some special cases it is even possible to determine a critical
Reynolds number by linearizing the governing equations. However, in
shear flows such as pipe, channel and Couette flows turbulence occurs
in experiments for sufficiently strong perturbations
\cite{reynolds1883,darbyshire1995,daviaud1992,hof2003} in parameter
regimes where the steady laminar flow is stable to infinitesimal
disturbances.

One of the striking features of linearly stable flows, first
recognized by Osborne Reynolds, is that disordered and smooth motion
can coexist \cite{reynolds1883}. Here turbulence is at onset
restricted to localized patches (called puffs in pipe flow) embedded
in a laminar background. Recently, streamwise-localized solutions of
the Navier--Stokes equations that share key spatial properties with
puffs but are time-periodic have been discovered \cite{avila2013}.
Although turbulent puffs are transient
\cite{hof2006,hof2008,avila2010,kuik2010}, spatial proliferation in
the form of splitting can balance the decay of individual puffs and
lead to an overall sustainment of turbulence
\cite{moxey2010,kavila2011}. The persistent turbulent state emerging
through this non-equilibrium phase transition, occurring at $\Rey_c
\simeq 2040$ \cite{kavila2011}, consists of a spatially-temporally
intermittent flow where individual turbulent clusters are transient
but proliferate faster than they decay. At the same time turbulence
can only be sustained locally, driven by the upstream laminar motion
\cite{hof2010}. At the downstream end the plug like velocity profile
needs to recover towards the laminar parabolic profile before a new
turbulent region can arise. This results in a minimum spacing of
turbulent clusters \cite{samanta2011} and an effective recovery length
\cite{barkley2011}. At higher \Rey\ puffs are superseded by
continuously growing structures called slugs \cite{wygnanski1973}, and
it is usually assumed that given sufficient time turbulence will fill
the entire pipe. This state depleted of laminar intermissions will in
the following be referred to as fully turbulent flow.

Althoguh laminar-turbulent intermittency was first reported in the
early experiments of Reynolds \cite{reynolds1883}, the probably first
quantitative characterization of intermittency in pipes was provided
by Rotta \cite{rotta1956}. In his experiments he continuously
disturbed the flow at the pipe inlet to generate turbulence and
monitored its evolution at different downstream distances. He
quantified the turbulent fraction and concluded that a state of fully
turbulent flow would be always reached above the critical Reynolds
number $\Rey_c$, which he speculated to be about $2000$
\cite{rotta1956}. Wygnanski and coworkers later recongized the
intermittent nature of turbulence in the transitional regime and
concluded that a fully turbulent flow would be realized at about
$\Rey\gtrsim 2700$ \cite{wygnanski1973}. More recently, the increase
in computing power has made it possible to numerically simulate
intermittent flows in spatially extended domains
\cite{willis2008,moxey2010,kavila2011}. One advantage of simulations
is that due to periodic boundary conditions in the axial direction the
dynamics of turbulent patches can be studied for very long times,
whereas in experiments they are convected out. Using simulations in
pipes of up to $125$ diameters in length, Moxey and Barkley
\cite{moxey2010} argue that no laminar islands are found in turbulent
flow beyond $\Rey\simeq 2600$. However, results from a model of pipe
flow, allowing for very long domains and observation times, suggest
that neither the transition to fully turbulent flow nor the transition
from puffs to slugs is sharp \cite{barkley2011}. In this paper we show
that laminar-turbulence intermittency is an intrinsic feature of shear
flow. As the Reynolds number increases laminar regions become scarcer
yet they do not disappear entirely. The underlying physical process
has its roots in a nearest neighbor interaction between turbulent
regions.

\section{Numerical simulations of long periodic pipes}

\begin{figure*}
  \centering
  \includegraphics[width=0.95\linewidth]{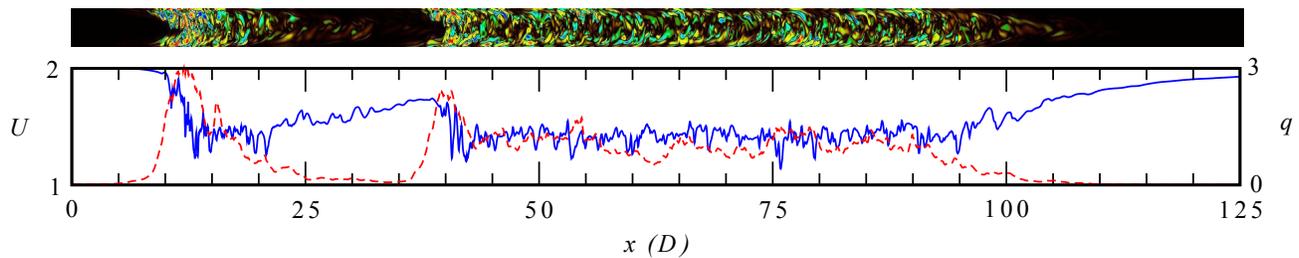}
  \caption{(Color online) Coexistence of a puff and a slug at
    $\Rey=2600$ in direct numerical simulations of pipe-flow. The flow
    is from left to right and $125$ diameters (D) of a $176D$-long
    pipe are shown. (Top) Snapshot of streamwise vorticity in a
    $(x,y)$ plane. (Bottom) Cross-sectionally averaged streamwise
    vorticity $q$ (dashed) and streamwise velocity $U$ (solid) along the
    pipe axis.}
  \label{slug2600}
\end{figure*}

A clear constraint of all previous investigations is the limited
observation times and pipe length accessible to numerical simulation
and laboratory experiment. In order to estimate the dynamical behavior
of pipe flow in the limit of infinite length and time, we performed
numerical simulations of periodic pipes of up to $500$ diameters in
length and considered observation times of up to $25,000D/U$ ($D$ is
the pipe diameter and $U$ the mean speed; the Reynolds number is
$\Rey=D U/\nu$). The Navier--Stokes equations were solved in
cylindrical coordinates $(r,\theta,x)$ using the hybrid spectral
finite-difference method of Willis \cite{willis2009}. The numerical
discretization consists of a non-equispaced $9$-point
finite-difference stencil in $r$ and of Fourier modes in $\theta$ and
$x$. The results for $\Rey\leq 2800$ have been obtained with up to
$N=64$ radial points and $M=\pm 36$ azimuthal Fourier modes, whereas
$K=\pm 1024$ axial Fourier modes have been used for a pipe length of
$32 \pi D \simeq 100D$. For $\Rey>2800$ we used $N=64$, $M=\pm 48$ and
$K=\pm 1280$.

The spreading of turbulence along the pipe was systematically studied
by using the following procedure. First, a puff was generated at low
Reynolds number, \eg $\Rey=2100$, from a localized
perturbation. Subsequently, the Reynolds number was impulsively
changed to a prescribed value and the dynamics was monitored in
time. Figure~\ref{slug2600}$a$ shows a snapshot of streamwise
vorticity in a $(x,y)$ plane at $t\simeq75D/U$ after the Reynolds
number was increased to $Re=2600$. Although initially the localized
puff appears to spread continuously, as a slug, it later splits into
two smaller turbulent structures, the first one resembling a slug and
the second one a puff.  Between them the cross-sectionally averaged
streamwise vorticity $q=\sqrt{\langle\omega_x^2\rangle_{r,\theta}}$
(red curve in Fig.~\ref{slug2600}$b$) rapidly approaches zero, whereas
the streamwise velocity at the centerline (blue curve) slowly
increases towards the laminar value $u=2U$, but does not quite reach
it. Our simulations confirm previous experimental observations, which
identified mixed occurrences of puffs and slugs originating from
single localized perturbations \cite{darbyshire1995,nishi2008}, and
suggest that the coexistence of localized and expanding structures
(\ie puffs and slugs) is intrinsic to pipe flow.

\section{Statistical analysis of laminar-turbulence intermittency}

\begin{figure}
  \centering
  \begin{tabular}{r}
      \includegraphics[width=0.88\linewidth,clip=]{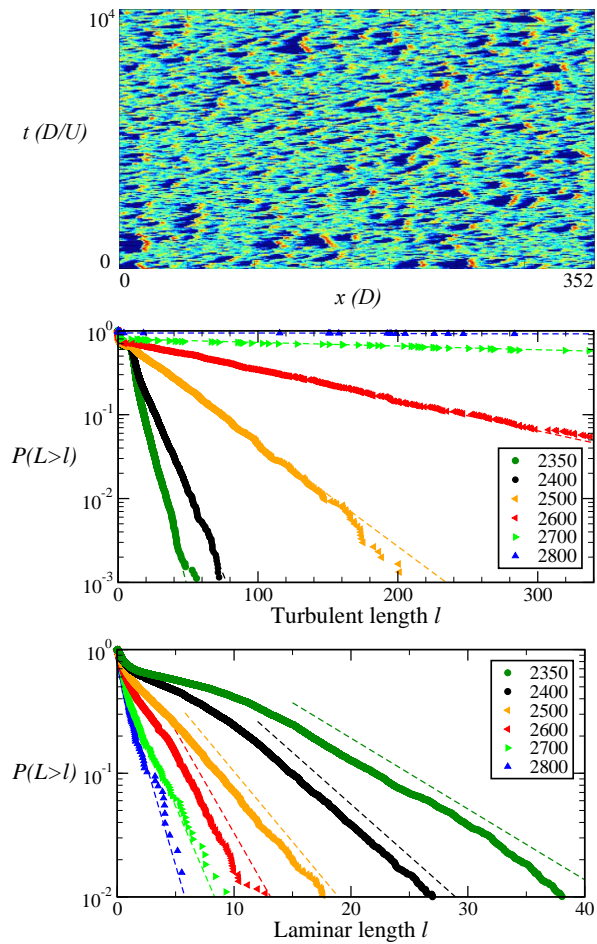}\\
      \includegraphics[width=0.9\linewidth,clip=]{fig2b.eps}\\
      \includegraphics[width=0.9\linewidth,clip=]{fig2c.eps}
  \end{tabular}
  \caption{(Color online) Top: Space-time diagram of
    cross-sectionally averaged streamwise vorticity $q$ in a frame
    co-moving at $0.93U$ at $Re=2500$. Dark blue corresponds to
    laminar flow and red to intense turbulence. Middle: Cumulative
    distribution functions of turbulent lengths at
    $\Rey\in[2350,2800]$ in a vertical log-scale. Bottom: Cumulative
    distribution functions of laminar lengths at $\Rey\in[2350,2800]$
    in a vertical log-scale.}
  \label{sizes}
\end{figure}

In order to study the asymptotic behavior of the system simulations
started from a localized disturbance were continued until the entire
domain was filled with spatio-temporal-intermittency. The flow
dynamics thereafter is shown in Figure~\ref{sizes}$a$, which is a
space-time diagram of $q(x,t)$ at $\Rey=2500$.  At a given instant in
time $t$, an $x$-constant cross section was considered laminar if
$q(x,t)<q^*=0.3U/D$. We note that our threshold choice yields a puff
size of $10D$ at $\Rey=2100$, which is consistent with results from
previous experimental studies considering the active part of the puff
\cite{hof2010}. Note that when streamwise velocity or pressure are
considered, the length of a puff is typically much longer due to the
slow recovery of the parabolic profile in the downstream direction
(Fig.~\ref{slug2600}$b$). Using this criterion the lengths of
turbulent segments $\Delta x$ were determined from each of the
snapshots in Fig.~\ref{sizes}$a$ and a collection of turbulent lengths
$l^i=\Delta x^i$ was generated. The cumulative distribution of $l^i$
is shown as left triangles in a logarithmic vertical axis in
Fig.~\ref{sizes}$b$ and is found to follow an exponential law for
lengths $l\gtrsim L^0_t=15D$ (patches longer than a puff). At low
Reynolds number ($\Rey\lesssim 2200$ at the time scales considered
here) spreading events (puff-splitting) are very rare. In the absence
of these events the length distributions are Gaussian-like about the
average puff length. At higher Reynolds number this behavior is still
observed for short patches $l\lesssim L^0_t$.

The curves in Fig.~\ref{sizes}$b$ show turbulent length distributions
in the range $\Rey\in[2350,2800]$. As the Reynolds number increases,
turbulent patches become increasinlgy longer because relaminarization
of pipe sections becomes more infrequent. The same analysis applied to
the laminar gaps between turbulent sections renders exponential
distributions as well (Fig.~\ref{sizes}$c$). Here also distributions
are exponential only for laminar gaps longer than the typical
interaction length at which puffs strongly influence each other
\cite{hof2010,samanta2011}. At $Re=2350$ this interaction length is
about $l\gtrsim L^0_l=15D$, whereas it rapidly decreases as $Re$
increases (at $\Rey\gtrsim 2700$, $L^0_l\lesssim 1D$). We note that
exponential distributions of the size of chaotic and laminar clusters
are a typical signature of spatio-temporal-intermittency and were
previously reported in simulations of Coupled Map Lattices
\cite{chate1988}, convection in an annulus \cite{ciliberto1988}, in
the Taylor-Dean system \cite{degen1996} and torsional Couette flow
\cite{cros2002}. Although the distribution of laminar lengths is
expected to be scale invariant (algebraic) at the onset of sustained
spatio-temporal-intermittency ($\Rey_c\simeq 2040$ in pipe flow), the
relevant physical decay and spreading processes occur at a time scale
$>10^7D/U$ \cite{kavila2011}, clearly out of reach in the
simulations. Using very long simulations of a reduced model of plane
Couette flow, Manneville \cite{manneville2009} has demonstrated that
laminar size distributions are algebraic at onset. More recently,
scale invariance at the onset of turbulence has been shown from
simualtions of the Navier--Stokes equations for Couette flow in a
tilted domain \cite{shi2013}. In pipe flow, reduced models that
recover the main features of the transitional dynamics have been also
proposed \cite{willis2009} and might prove useful in addressing this
point.  Here we checked that the laminar size distributions are indeed
exponential for $\Rey\gtrsim2350$, whereas we found that for
$\Rey\lesssim 2300$ our simulations cannot be used to quantiatively
estimate the distributions. Exploring this low $\Rey$ regime remains a
challenge and would require time-integrations and domains
substantially beyond those used in this work.

\begin{figure}
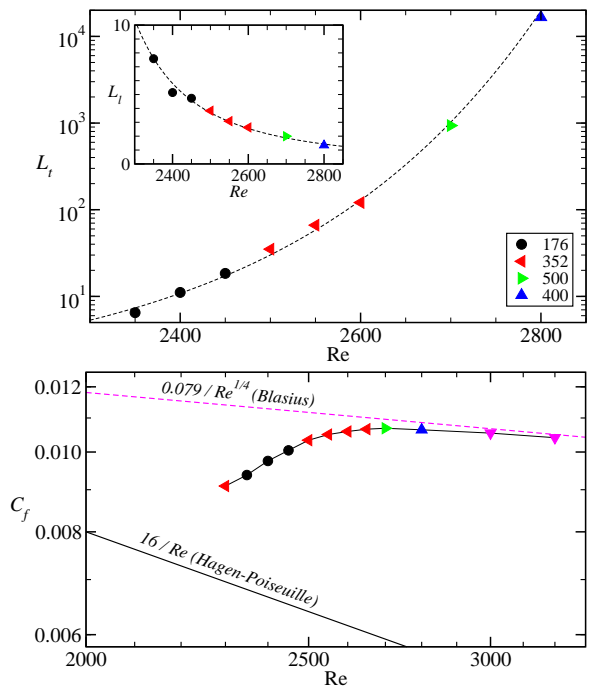

  \centering
  \begin{tabular}{r}
    \includegraphics[width=0.85\linewidth,clip=]{fig3a.eps}\\
    \includegraphics[width=0.89\linewidth,clip=]{fig3b.eps}
  \end{tabular}
  \caption{(Color online). Top: Characteristic turbulent size,
    corresponding to the slope of the dashed lines in
    Fig.~\ref{sizes}$c$, as a function of $\Rey$. The computational
    domain length is indicated in the legend. Inset: Characteristic
    laminar gap length as function of $\Rey$. Bottom: Skin friction
    coefficient $C_f=-\langle \partial_z p\rangle_{r\theta z
      t}D/(2\rho U^2)$, where $\rho$ is the fluid density.}
  \label{meansize}
\end{figure}

The distributions of the form $\exp[(l-L^0_t)/L_t]$ in
Fig.~\ref{sizes}$b$ naturally define a characteristic turbulent length
$L_t$ at each Reynolds number. The variation of the scale parameter
$L_t$ as a function of $\Rey$ is shown in a logarithmic vertical axis
in Fig.~\ref{meansize}$a$. As $\Rey$ increases $L_t$ grows strikingly
fast (Fig.~\ref{meansize}$a$). The data are best approximated by a
super-exponential fit of the shape $L_t=\exp[\exp(a + b\,\Rey)]$ and
indicate, if extrapolated, that no divergence of turbulent lengths
occurs at a finite \Rey.  This result suggests that no continuously
growing slugs exist, but rather these split into smaller slugs after
perhaps a very long, but finite time.  Moreover, the fit approximates
the data well into the puff regime and hence provides a quantitative
link between localized and pipe-filling turbulent flow.  In the case
of laminar gaps it was found that the analogous scale parameter $L_l$
decreases algebraically with Reynolds number (see the inset in
Fig.~\ref{meansize}$a$), also supporting that fully turbulent flow is
reached only in the asympotic \Rey\ limit. Note that $L_l$ is not the
average laminar gap length because it does not contain information
about the part of the distribution which is not exponential. For
example $L_l\sim 7.5D$ at $\Rey=2350$ but the average laminar gap
length is $L_l+L^0_l=22.5D$.

We verified that the length distributions are independent of initial
conditions by starting simulations with turbulent flow at large
Reynolds number and quenching to the desired \Rey. These lead to the
same size distributions as the simulations started from a single
puff. Secondly, we repeated the simulations for different pipe
lengths. Again convergence to the same distributions was
achieved. Overall, the results show that the asymptotic size
distributions are independent of the pipe length and initial
conditions, and so they are an intrinsic property of the system in the
`thermodynamic limit'.  Although lengths (durations) of laminar and
turbulent regions in pipe flow had been previously investigated in
experiments \cite{rotta1956,wygnanski1973,sreenivasan1986}, the
observation times of $<500D$ where too short to reach size
distributions which are in statistical equilibrium (see especially
Fig.~26 of Ref.~\onlinecite{rotta1956}). Hence, the lengths of the
observed laminar episodes was likely controlled by the time-scale at
which turbulence was spontaneously triggered by the disturbance
(obstacle) at the inlet.

While a variation of the cut-off $q^*$ discriminating laminar from
turbulent flow necessarily leads to a shift of absolute values, the
qualitative scaling remains unchanged. Regardless of the choice, the
characteristic turbulent (laminar) size scale super-exponentially
(algebraically) with \Rey. It was found that turbulent distributions
are more sensitive to the threshold value than laminar ones, which
remain almost unchanged. 

The gradual transition from spatio-temporal-intermittency to fully
turbulent flow is further illustrated in Fig.~\ref{meansize}$b$,
showing the skin friction dependence on \Rey. In contrast to previous
simulations \cite{willis2008} we do not observe an overshoot of the
Blasius curve as this is approached. This discrepancy is likely due to
numerical resolution. It is noted that to accurately resolve friction
values high numerical resolutions are required. Here convergence was
checked by independently increasing the resolution in all directions
until the difference between computed frictions was below
$0.5\%$. Beyond $\Rey>3000$ the flow rapidly converges to the Blasius
friction law, whereas for $\Rey\lesssim 2300$ it was not possible to
reliably compute friction values due to the extremely slow time-scales
at which puffs merge, split and annihilate each other to open laminar
gaps \cite{kavila2011}.

\section{Transient laminar islands in turbulent flow}

\begin{figure}
  \centering
  \includegraphics[width=0.98\linewidth]{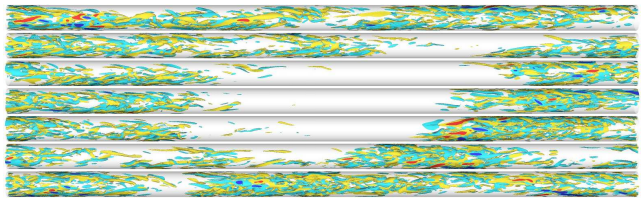}
  \caption{(Color online). Streamwise vorticity isosurfaces showing
    the development of a laminar island in turbulent flow at
    $\Rey=2800$. The time-snapshots are separated by $28D/U$ in a
    frame co-moving at the mean-speed $U$. Here $23D$ out of a domain
    of $400D$ are shown.}
  \label{lamisland}
\end{figure}

The development of a laminar island at $\Rey=2800$ is shown in
Fig.~\ref{lamisland} and illustrates the collapse of streamwise
vortices in an extended part of the flow domain. A similar phenomenon
was observed in minimal channels, where occasionally streamwise
vortices weaken across the entire domain and turbulent activity
temporarily ceases \cite{xi2010}. Our observation that laminar islands
keep emerging in the flow as the Reynolds number is increased supports
that quiescent regions are intrinsic to turbulent shear flow in
realistically long domains. Unlike in Couette flows, where
laminar-turbulent patterns have been suggested to emerge as a
wavelength instability of the fully turbulent flow
\cite{prigent2002,barkley2005,manneville2012}, in pipe flow laminar
domains are found to appear at random locations and times in the
turbulent flow (see Fig.~\ref{sizes}a). Our results are in agreement
with recent experiments in long pipes \cite{samanta2011} which found
random spacing between puffs following Reynolds number reductions from
fully turbulent flow. The essence of this process is captured by
excitable and bistable media models \cite{barkley2011}: the emergence
of a laminar gap depends here only on the state of turbulence in the
nearby region.  The results presented here can be used to test and
calaibrate models of pipe flow that have been recently developed
following different approaches
\cite{barkley2011,sipos2011,allhoff2012}. The exponential size
distributions of laminar and turbulent lengths further suggest that
laminar turbulent patterns in pipe flow are the manifestation of a
contact process only asymptotically giving rise to fully turbulent
flow.

\section{Discussion}

The traditional view of equilibrium puffs giving way to a regime of
puff splitting and eventually to one of expanding slugs
\cite{wygnanski1973,wygnanski1975} has originated from observations
over time-scales typically accessible in laboratory experiments. While
over these timescales this picture appears to reflect the sequence of
events, it nevertheless obscures some of the most relevant
physics. Only the observation of a large number of events in extended
domains and over long times reveals that a uniform state of turbulence
(or fully turbulent flow) does not exist. Instead strong
spatio-temporal fluctuations are intrinsic to the flow including the
extreme case where turbulence collapses in some part of the domain. As
we have pointed out elsewhere \cite{kavila2011}, localized turbulent
patches are also never in equilibrium, \ie they either grow or decay,
and an equilibrium puff regime does not exist either.

In conclusion, only one state of turbulence exists which is that of a
spatio-temporally intermittent flow exhibiting large fluctuations. At
low $\Rey>Re_c\approx 2040$ fluctuations manifest themselves in
sequences of laminar and turbulent regions, whereas at $\Rey> 3000$
laminar events become so scarce that turbulence flow will here appear
as space filling but with large fluctuations of intensity in space and
time providing the familiar turbulence intermittency still observed at
Reynolds numbers many orders of magnitude larger.

\begin{acknowledgments} 
  We thank A.~P.~Willis for sharing his hybrid spectral
  finite-difference code. We thank K.~Avila for correcting our
  interpretation of Rotta's paper in an earlier draft. We are grateful
  to D.~Barkley, N.~Goldenfeld, M.~Holzner, A.~de~Lozar and
  A.~P.~Willis for dicussions.  The research leading to these results
  has received funding from the Max Planck Society and the European
  Research Council under the European Union's Seventh Framework
  Programme (FP/2007-2013) / ERC Grant Agreement 306589. We
  acknowledge computing resources from GWDG and the J\"ulich
  Supercomputing Centre (grant HGU16).
\end{acknowledgments}

\end{document}